7-22-2021

# Digital Literacy and Reading Habits of the Central University of Tamil Nadu Students: A Survey Study

Subaveerapandiyan A
*Regional Institute of Education Mysore*, subaveerapandiyan@gmail.com

Priyanka Sinha
*Punjab University, Chandigarh*, priyankasinha101099@gmail.com

# Digital Literacy and Reading Habits of the Central University of Tamil Nadu Students: A Survey Study

**Subaveerapandiyan A**
Professional Assistant
Regional Institute of Education Mysore, India
Email: subaveerapandiyan@gmail.com
**Priyanka Sinha**
PhD. Research Scholar
Punjab University, Chandigarh
Email: priyankasinha101099@gmail.com

**Abstract**
*The study attempted to understand the University students' digital reading habits and their related skills. It also has a view of students' preferred sources of reading, whether physical or digital resources. For this study, we conducted a survey study with students and research scholars of the Central University of Tamil Nadu, India. The instrument was a structured questionnaire distributed with various modes. The result found that the majority of the students are well known about digital tools and usage, most of the students are excellent in digital literacy skills and other findings is however they are good in digital literacy even though they like to read print books is their most favorable preference. The results conclude that whatever technological devices are developed and students have also grown their technical knowledge. The result finds out, in education especially reading-wise, students or readers' first wish is printed resources; digital books are secondary to them.*

1. **Introduction**

Digital literacy plays a vital role in day today life such as the workplace environment, educational institutes, all kinds of organizations that depend on digital. In higher educational institutes, libraries provide their services, most probably digital based and classes also conducted online. For that literacy skill acquiring is an essential especially in digital literacy skills. Literacy skills used for information seeking and digital literacy helps to know, evaluate, capture, and measure the sources of existing knowledge. Skills and knowledge are different from one to another, various sources used for acquiring the information and knowledge. Sufficient technological and related skills can make a good academic scholar.

Due to exponential growth in population and globalization, there was immense influence over communication and mobilization of information. The digital world has created a range of opportunities to access information remotely globally and address a knowledge gap. Digital literacy emerged as a competency to understand the information needed digitally, rectify the best possible source, and evaluate its authenticity and communicate that information. Digital literacy comprises the skills of media, computer and internet literacy as well. (**Ayhan 2016**) Spires and Bartlett discussed the three fundamental skills of digital literacy, i.e., locating and consuming digital content, creating digital content, and communicating digital content. (**Khosrow-Pour 2017**) This paper tries to focus on the reading habit of students in the digital niche, accessing their ability to fetch information from various sources remotely in pandemic situations.

2. **The objective of the Study**
   - ❖ To know the student's digital literacy skills
   - ❖ To find out the student reading habits
   - ❖ To understand the student's preference source for reading electronic or print
   - ❖ To know about student's digital application and software skills

3. **Review of Literature**

   *Digital literacy indicator*

   **Techataweewan** tried to discover the concept of digital literacy in Thai society, and digital literacy indicators were also discussed using confirmatory factor analysis. Four major factors behind digital literacy indicators like operation skills which include cognitive, inventive, and presentation skills. Thinking skills include analyzing, evaluating, and creativity of undergraduate students. Collaboration has an essential role in digital literacy, as it assists in filling the knowledge gap by extracting information from various sources and making user information independent. It requires teamwork, sharing of data, and work in a network. A digitally literate person should also be aware of ethical and legal hunches in data sharing and management. (**Techataweewan 2018**)

   *Digital literacy skills*

   **Johnston** tried to explore the digital literacy framework in Australian libraries to equip students with digital literacy skills and suggested digital literacy practice among library schools. The foreseen steps required for integration of digital literacy skills are the formulation of an advisory committee, students being the part of this committee, consistent approach towards digital literacy skills, mapping course learning outcomes, and documenting and collaborating digital literacy initiatives at the university level. (**Johnston 2020**)

   The best way to figure out if the person is digital literate or not is by assessing their digital literacy skills, competency to use information resources in driving information, and tools to evaluate that information. **Shwetha K** explored the digital information literacy skills among faculty members in engineering college in Mangalore. The author used various parameters like frequency of using the internet, using information resources, and multiple sources consulted for fetching information. The faculty members were excellent at using web resources with accuracy. (**Shwetha K 2017**)

**Iqbal Singh** tried to illustrate digital literacy skills among healthcare professionals at GGS Medical College, Faridkot, Punjab. Healthcare professionals are the first line of defense in a pandemic, so they should be proficient enough to deal with raw data for further research in the healthcare sector. The majority of respondents in this study, i.e., 84%, were aware of internet applications like MS office. 94 % of them were using data from various e-resources in research work, and 80 % were able to judge the authenticity and reliability of that information. (Iqbal **Singh 2015**)

**Kaeophanuek** surveyed Thai students to know about students' digital literacy skills and the environment likely to be provided for digital literacy skills. Information professionals need to be digital specialists; for this, they need to have basic digital data management skills, use digital tools and cognitively create content. Digital information usage policies are like a blueprint while using digital content by university students. To nurture their digital reading habits, proper infrastructure and instruction are required at the digital level. (**Kaeophanuek 2018**)

**Khatun** surveyed public library professionals in Norway to explore digital literacy skills and find out the barriers to improving those skills. Library professionals suggest three barriers in improving digital literacy skills, i.e., organizational barriers, Personal barriers, and Technological barriers. Experience also plays a vital role in improving digital literacy skills, so sharing information by experienced library experts with young professionals can address this barrier. Training and regular orientation programs help in technological obstacles. (**Khatun 2015**)

**Anjaiah** conducted an exploratory study at Dravidian University to assess the digital literacy skills of research scholars and students. The study report revealed that most of the respondents were using the internet, and smartphones were the means to access information they were using daily for browsing e-books. The maximum number of students were satisfied with the digital information resources. Conventional computer literacy skills are significant to complement digital literacy skills that are just not limited to digital devices. (**Anjaiah 2016**)

**Jeffrey** conducted a case study design among four higher education institutions to explore the obstacles and support required by students in developing digital information literacy. Competency development at the digital level is not simply exposure to technology, but skill development protocols need to be followed. Significant hindrances to this process are socio-economic barriers, gender bias, age gap, and acceptance of new technology. Collaborative learning with the use of social media is a potential solution to address this barrier. (**Jeffrey 2011**)

**Parvathamma N.** conducted the study among the student community in management institutes in Davanagere District, Karnataka, to understand the ICT tools and web-based services used by students to frame the curriculum for digital literacy courses. The study revealed that most of the respondents own their personal computers with internet connectivity. Students were preferably using laptops for classwork. The email was the top web 2.0 tool used by students for personal uses. Students were aware of the literacy and ICT tools but did not make proper use of them, so proper professional training was suggested to impart independent digital users. (**Parvathamma N. 2013**)

**Emiri** tried to explore contemporary digital literacy skills among Librarians In University Libraries in Edo and Delta States, Nigeria. Most of the librarians were using Email for communication, they acquired digital literacy skills through IT programs, but they were using it at a moderate level. Digital literacy skills have shown a positive impact on the delivery of library services. Barriers in delivering digital skills were lack of digital facility, fund constraints, and lack of training. Libraries require competency development programs or digital literate librarians should be recruited. (**Emiri 2015**)

*Use of digital literacy*
**McDougall** conducted a project on digital literacy skills among students of the age group of 6-9 on digital classrooms and community space usage. Community stakeholders had limited access to mobile literacy tools, limited skills, and technology barriers. Lack of funds, time constraints, anxiety around screen time amplified the negative outcome of school pedagogy. (**McDougall 2018**)

4. **Central University of Tamil Nadu**
The Central University of Tamil Nadu was recently established in the year of 2009. Presently, universities have 12 schools, 27 academic departments and 160 teaching faculties (2021 website data). They were offering Undergraduate, Postgraduate, Integrated UG and PG, PG Diploma and Research programmes. It's a coeducational university. The university Central Library has more than 36500 books and subscribed 130 various discipline print journals and 2187 eBooks. The library is open for users the whole week with different timings.

5. **Scope, Limitation and Methodology**
The research design used for this present study is quantitative design. The sample for the current study and include limitation of the study is Central University of Tamil Nadu Undergraduate and Postgraduate students as well as research scholars. The data sample consists of 135 students and scholars from the Central University of Tamil Nadu, India (Male 28.9% and 71.7%). The questionnaire was issued through the random sampling method. For collecting information on digital literacy and reading habits, a variety of digital tools and techniques were used. This study we used a survey method and the structured-questionnaire was distributed among the students through their official institute Email ID, WhatsApp, and Telegram. The data is mainly regarding their digital literacy skills, digital devices and tools, application software skills, reasons for reading, etc. This study used the Likert 5-point scale.

## 6. Data Analysis and Interpretations

**Table 1: Demographic Frequency Distribution of Respondents**

| Type | Division | Frequency | Percentage |
|---|---|---|---|
| Gender | Male | 39 | 28.9 |
| | Female | 96 | 71.1 |
| Age Groups (In years) | 17-21 | 66 | 48.9 |
| | 22-27 | 61 | 45.2 |
| | 28-35 | 6 | 4.4 |
| | 35-50 | 2 | 1.5 |
| | 50 and above | 0 | 0 |
| Location | Urban | 39 | 28.9 |
| | Semi-Urban | 26 | 19.3 |
| | Rural | 70 | 51.9 |
| Current School of Study | Basic & Applied Sciences | 8 | 5.9 |
| | Mathematics & Computer Sciences | 28 | 20.7 |
| | Social Sciences & Humanities | 20 | 14.8 |
| | Behavioral Sciences | 0 | 0 |
| | Commerce & Business Management | 23 | 17 |
| | Communication | 27 | 20 |
| | Education & Training | 8 | 6 |
| | Technology | 10 | 7.4 |
| | Performing Arts & Fine Arts | 4 | 3 |
| | Earth Sciences | 2 | 1.5 |
| | Life Sciences | 5 | 3.7 |
| | Legal Studies | 0 | 0 |
| Current Educational Status | Undergraduate | 11 | 8.1 |
| | Postgraduate | 75 | 55.6 |
|  | Integrated UG/PG | 35 | 25.9 |
|  | Research Scholar (M.Phil./PhD.) | 14 | 10.4 |
| **Total** |  | **135** | **100** |

Table 1 shows the demographic distribution of the respondents. **Gender** wise 71.1% female respondents and 28.9% male respondents, its shows female respondents are high; **Age** group-wise respondents 48.9% respondents are 17 to 21 years old, 45.2% respondents are 22 to 27 years, 4.4% respondents are 28 to 35 years and 1.5% respondents are 35 to 50 years; **Location** wise shows that 28.9% of respondents are urban, 19.2% respondents are semi-urban and 51.9% respondents are rural; **Current school of study** depicts the highest 20.7% respondents in Mathematics and Computer Science, followed by 20% respondents are communication and the least respondents 1.5% are Earth Sciences and 0% respondents are from Behavioral Sciences and Legal Studies; **Current educational status** wise 8.1% respondents are Undergraduate, 55.6% respondents are Postgraduate, 25.9% respondents are Integrated Postgraduate and 10.4% respondents are Research Scholars.

**Figure 1: Sources for Knows About New Technologies**

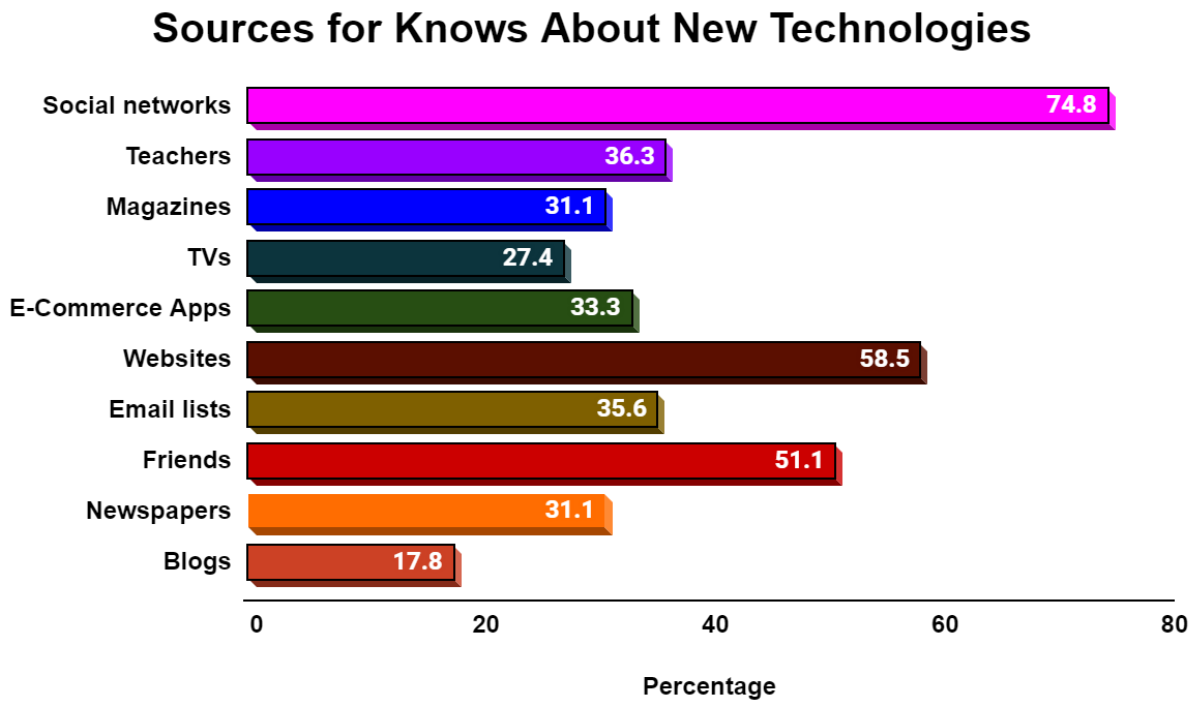

Figure 1 reveals students using sources for knowing about new technology; it depicts 74.8% of respondents source is social networks, followed by 58.5% respondents used websites and 51.1% respondents learned from friends.

Table 2: Self-Rating of Digital Literacy Skills

| Digital Literacy Skills | Very Good | Good | Acceptable | Poor | Very Poor | Mean | SD |
|---|---|---|---|---|---|---|---|
| Typing skills | 38 (28.1%) | 58 (43%) | 32 (23.7%) | 4 (3%) | 3 (2.2%) | 3.91 | 0.91 |
| Web Search Skills | 41 (30.4%) | 61 (45.2%) | 28 (20.7%) | 3 (2.2%) | 2 (1.5%) | 4 | 0.85 |
| Computer Literacy | 35 (25.9%) | 67 (49.6%) | 31 (23%) | 0 (0%) | 2 (1.5%) | 3.98 | 0.79 |
| Internet literacy | 34 (25.1%) | 75 (55.6%) | 24 (17.8%) | 0 (0%) | 2 (1.5%) | 4.02 | 0.75 |
| Digital Literacy | 30 (22.3%) | 61 (45.2%) | 37 (27.4%) | 4 (3%) | 3 (2.2%) | 3.82 | 0.86 |

**Scale Used: Very Good=5, Good=4, Acceptable=3, Poor=2, Very Poor=1, SD= Standard Deviation**

Figure 2: Self-Rating of Digital Literacy Skills

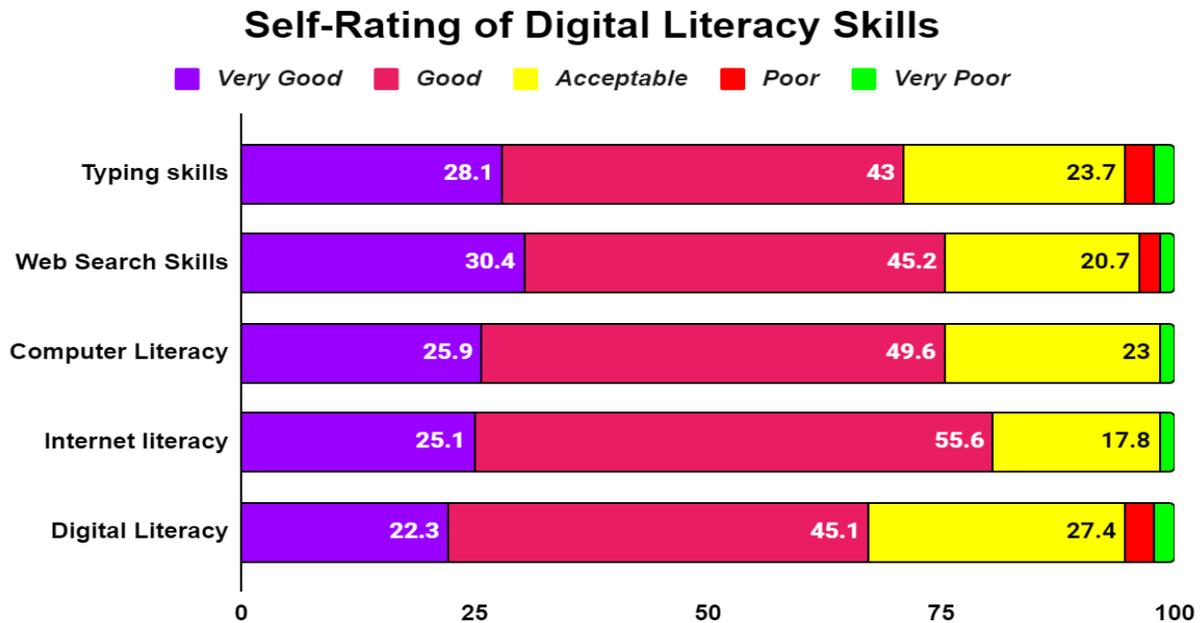

Table 2 reported self-rating of digital literacy skills. Majority of the students respondents 43% are good and 28.1% very good in typing skill; Web search skills 45.2% of respondents are good and 30.4% of respondents are very good in searching; Computer literacy skill 49.6% of respondents are good and 25.9% respondents were very good; Internet literacy skills one-half of the respondents 55.6% are good and 25.1% respondents are very good; Digital literacy skills the ability of using digital technologies 45.2% good and 22.3% respondents are very good in digital literacy skills.

**Table 3: Digital Literacy Skills**

| Digital Literacy Skills | Yes | No |
|---|---|---|
| Understand the basic functions of computer hardware components | 118 (87.4%) | 17 (12.6%) |
| Do you use keyboard shortcuts? | 118 (87.4%) | 17 (12.6%) |
| Do you use the computer for learning purposes? | 120 (88.9%) | 15 (11.1%) |
| Do you use social networking services? | 112 (83%) | 23 (17%) |
| Do you have mobile apps you use for language learning purposes? | 97 (71.9%) | 38 (28.1%) |
| Can you create and update web pages? | 52 (38.5%) | 83 (61.5) |

**Figure 3: Digital Literacy Skills**

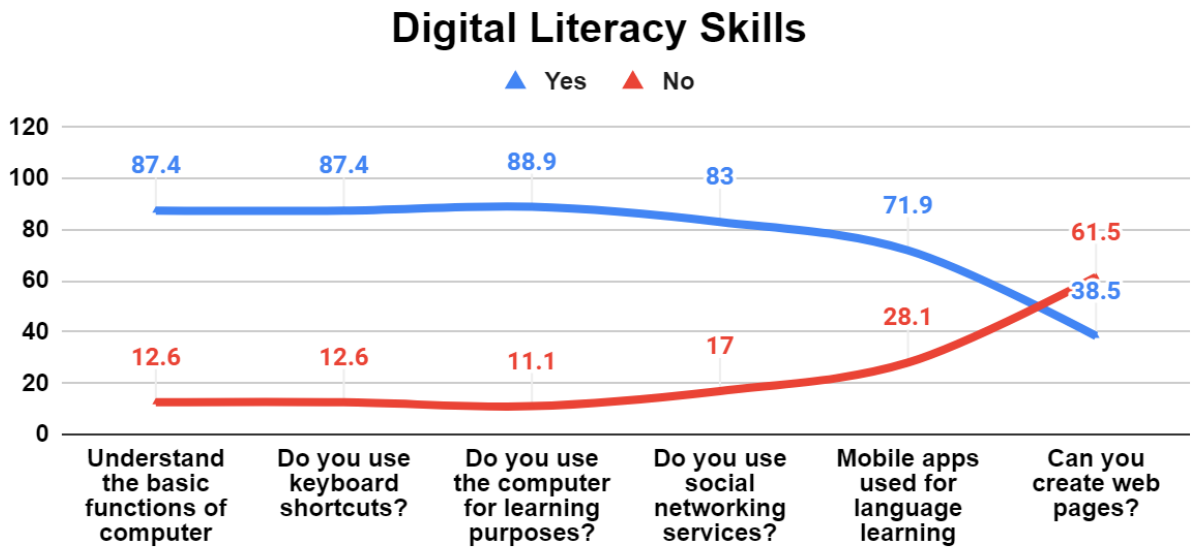

The findings in table and figure 3 show digital literacy skills. Understanding the basic functions of computer hardware components majority 87.4% respondents say 'yes' and 12.6% of respondents say 'no'; Knowledge of using keyboard shortcuts 87.4% respondents marked 'yes' and 12.6% respondents marked 'no'; 88.9% most of the respondents using the computer for learning purpose and 11.1% used for multipurpose; 83% vast percentage of respondents using the social networking services and 17% least respondents not used; the highest 71.9% of respondents used language learning mobile apps and 28.1% respondents did not use; 38.5% respondents aware of creating and update web pages and 61.5% respondents unaware of creating web pages.

Table 4: Frequency of Using Digital Environment

| Frequency of Using Digital Environment | Very Frequently | Frequently | Occasionally | Rarely | Never | Mean | SD |
|---|---|---|---|---|---|---|---|
| Word processor | 55 (40.7%) | 40 (29.6%) | 18 (13.4%) | 19 (14.1%) | 3 (2.2%) | 3.92 | 1.14 |
| Email | 81 (60%) | 40 (29.6%) | 5 (3.7%) | 7 (5.2%) | 2 (1.5%) | 4.41 | 0.9 |
| World Wide Web | 68 (50.4%) | 38 (28.1%) | 13 (9.6%) | 11 (8.2%) | 5 (3.7%) | 4.13 | 1.11 |
| Database | 29 (21.4%) | 43 (31.9%) | 25 (18.5%) | 26 (19.3%) | 12 (8.9%) | 3.37 | 1.26 |
| Spreadsheet | 34 (25.2%) | 39 (28.9%) | 28 (20.7%) | 27 (20%) | 7 (5.2%) | 3.48 | 1.21 |
| Language App | 42 (31.1%) | 34 (25.1%) | 26 (19.3%) | 26 (19.3%) | 7 (5.2%) | 3.57 | 1.25 |
| Blog | 17 (12.6%) | 33 (24.4%) | 29 (21.5%) | 32 (23.7%) | 24 (17.8%) | 2.9 | 1.3 |
| Text chatting | 89 (66%) | 32 (23.7%) | 6 (4.4%) | 5 (3.7%) | 3 (2.2%) | 4.49 | 0.9 |
| Voice chatting | 74 (54.8%) | 31 (23%) | 16 (11.9%) | 13 (9.6%) | 1 (0.7%) | 4.21 | 1.03 |
| Video conferencing | 54 (40%) | 43 (31.9%) | 20 (14.8%) | 14 (10.3) | 4 (3%) | 3.95 | 1.11 |
| Electronic dictionary | 49 (36.3) | 45 (33.3%) | 26 (19.3%) | 10 (7.4%) | 5 (3.7%) | 3.91 | 1.08 |

**Scale Used: Very frequently=5, Frequently=4, Occasionally=3, Rarely=2, Never=1, SD= Standard Deviation**

Figure 4: Frequency of Using Digital Environment

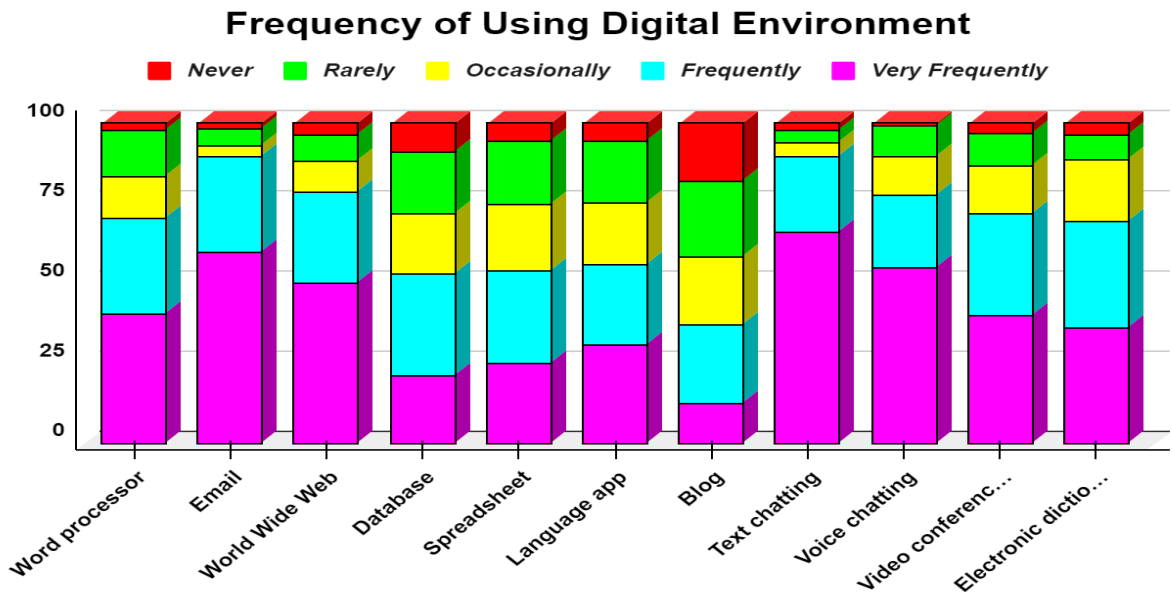

In table 4 and figure 4, the respondents were asked about the frequency of using the digital environment. The majority of the respondents 40.7% very frequently used a word processor, 60% of respondents very frequently used email, 50.4% of respondents were very frequently used world wide web, 31.9% of respondents used database frequently, 28.9% of respondents were used spreadsheet frequently, 31.1% respondents used language learning mobile apps very frequently, 24.4% of respondents were used blog frequently, 66% of respondents were used text chatting very frequently, 54.8% respondents were preferred voice chatting very frequently, 40% respondents very frequently used video conferencing very frequently, 36.3% of respondents were used very frequently electronic dictionary.

**Table 5: Self Rate of Digital Application Skills**

| Digital Application Skills | Very Good | Good | Acceptable | Poor | Do not know | Mean | SD |
|---|---|---|---|---|---|---|---|
| Word processing | 48 (35.6%) | 45 (33.4%) | 40 (29.6%) | 1 (0.7%) | 1 (0.7%) | 4.02 | 0.86 |
| Spreadsheet | 29 (21.5%) | 24 (17.8%) | 65 (48.2%) | 14 (10.3) | 3 (2.2%) | 3.67 | 0.99 |
| Database | 18 (13.4%) | 14 (10.3%) | 53 (39.3%) | 30 (22.2%) | 20 (14.8%) | 2.85 | 1.2 |
| Presentation | 40 (29.6%) | 37 (27.4%) | 48 (35.6%) | 7 (5.2%) | 3 (2.2%) | 3.77 | 1 |
| Communication | 30 (22.2%) | 21 (15.6%) | 52 (38.5%) | 20 (14.8%) | 12 (8.9%) | 3.27 | 1.21 |
| Social networking | 30 (22.2%) | 37 (27.4%) | 49 (36.3%) | 10 (7.4%) | 9 (6.7%) | 3.51 | 1.11 |
| Search engines | 60 (44.4%) | 39 (28.9%) | 31 (23%) | 3 (2.2%) | 2 (1.5%) | 4.12 | 0.94 |

**Scale Used: Very Good=5, Good=4, Acceptable=3, Poor=2, Don't Know=1, SD= Standard Deviation**

**Figure 5. Self-Rate of Digital Application Skills**

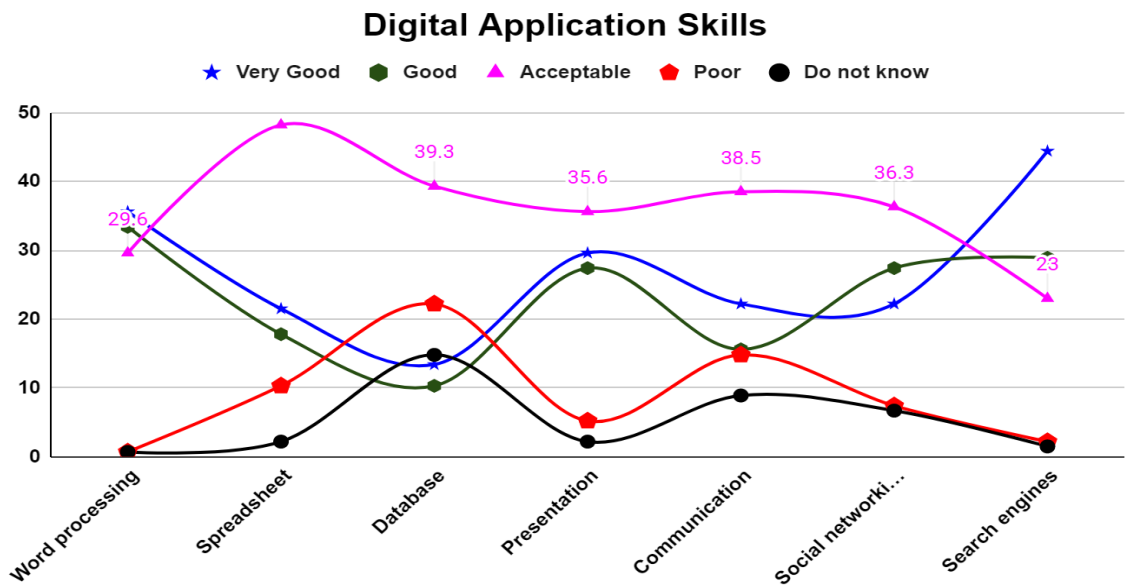

As seen in table 4 and figure 5 self-rate on digital application skills. In this table 35.6% the most of the respondents very good in word processing applications, 48.2% respondents acceptable in spreadsheet applications, 39.3% of respondents acceptable in database applications, 35.5% respondents acceptable in presentation applications, 38.5 respondents acceptable in communication applications, 36.3% respondents acceptable in social networking services, 44.4% respondents very good in the usage of web search engines.

**Table 6: Digital Devices Usage**

| Digital Devices Usage | Strongly Agree | Agree | Uncertain | Disagree | Strongly Disagree | Mean | SD |
|---|---|---|---|---|---|---|---|
| I enjoy using digital devices | 60 (44.4%) | 58 (43%) | 11 (8.2%) | 6 (4.4%) | 0 (0%) | 4.27 | 0.79 |
| I feel comfortable using digital devices | 48 (35.5%) | 63 (46.7%) | 11 (8.2%) | 12 (8.9%) | 1 (0.7%) | 4.07 | 0.92 |
| I am aware of various types of digital devices | 44 (32.6%) | 64 (47.4%) | 19 (14.1%) | 8 (5.9%) | 0 (0%) | 4.06 | 0.83 |
| I understand what digital literacy is | 47 (34.8%) | 65 (48.1%) | 14 (10.4%) | 9 (6.7) | 0 (0%) | 4.11 | 0.84 |
| I am willing to learn more about digital technologies. | 65 (48.1%) | 51 (37.8%) | 14 (10.4%) | 3 (2.2%) | 2 (1.5%) | 4.28 | 0.85 |
| I think that it is important for me to improve my digital fluency. | 61 (45.2%) | 51 (37.8%) | 15 (11.1%) | 6 (4.4%) | 2 (1.5%) | 4.2 | 0.91 |
| I think that my learning can be enhanced by using digital tools and resources. | 56 (41.4%) | 63 (46.7%) | 11 (8.2%) | 4 (3%) | 1 (0.7%) | 4.25 | 0.78 |
| I think that training in technology-enhanced language learning should be included in language education programs. | 64 (47.4%) | 59 (43.7%) | 8 (5.9%) | 2 (1.5%) | 2 (1.5%) | 4.34 | 0.78 |

**Scale used: SA=5, A=4, Uncertain=3, D=2, Strongly Disagree=1, SD=Standard Deviation**

Table 6 shows digital device usage and their enjoyment. The 44.4% majority of respondents strongly agreed that they enjoyed using digital devices, 46.7% of respondents agreed they were comfortable using digital devices, 47.4% respondents agreed they were aware of various digital devices, 48.1% of respondents agreed knew about digital; literacy, and skills, 48.1% respondents strongly agreed they are willing to learn about digital technologies, 45.2% respondents strongly agreed digital fluency is important to improve themselves, 46.7% of respondents agreed their digital learning can be enhanced by using digital tools and digital resources, 47.4% of respondents strongly agreed they think that training in technology-enhanced language learning should be included in language education programs.

**Table 7. Knowledge about digital tools**

| Knowledge About Digital Tools | Strongly Agree | Agree | Neutral | Disagree | Strongly Disagree | Mean | SD |
|---|---|---|---|---|---|---|---|
| I know how to use digital tools to find information | 55 (40.7%) | 58 (43%) | 16 (11.8%) | 4 (3%) | 2 (1.5%) | 4.18 | 0.86 |
| know how to use digital tools to understand information | 56 (41.5%) | 61 (45.2%) | 13 (9.6%) | 4 (3%) | 1 (0.7%) | 4.23 | 0.8 |
| know how to use digital tools to connect with others | 54 (40%) | 64 (47.4%) | 12 (8.9%) | 3 (2.2%) | 2 (1.5%) | 4.22 | 0.81 |
| know how to use digital tools to work with others | 44 (32.6%) | 55 (40.7%) | 24 (17.8%) | 8 (5.9%) | 4 (3%) | 3.94 | 1 |
| I know how to use digital tools to create my work | 43 (31.9%) | 56 (41.5%) | 23 (17%) | 7 (5.2%) | 6 (4.4%) | 3.91 | 1.04 |
| I know how to use digital tools to share my work | 43 (31.9%) | 61 (45.2%) | 24 (17.7%) | 4 (3%) | 3 (2.2%) | 4.01 | 0.9 |
| I understand what it means to be a responsible digital citizen | 42 (31.1%) | 58 (43%) | 23 (17%) | 10 (7.4%) | 2 (1.5%) | 3.94 | 0.95 |
| I like learning while using digital tools | 54 (40%) | 56 (41.4%) | 19 (14.1%) | 2 (1.5%) | 4 (3%) | 4.14 | 0.92 |

**Scale used: SA=5, A=4, Neutral=3, D=2, Strongly Disagree=1, SD=Standard Deviation**

Table 7 indicates knowledge about digital tools. 43% the highest percentage of respondents agreed and followed by 40.7% respondents strongly agreed to know how to use digital tools to find information; 45.2% respondent agreed and 41.5% respondents are strongly agreed know how to use digital tools to understand information; 47.4% of respondents agreed and 40% of respondents strongly agreed to know how to use digital tools to connect with others; 40.7% of respondents agreed and 32.6% of respondents are strongly agreed know how to use digital tools to work with others, 41.5% respondents agreed and 31.9% respondents strongly agreed to know how to use digital tools to create my work; 45.2% respondents agreed and 31.9% of respondents strongly agreed to know how to use digital tools to share my work; 43% of respondents agreed and 31.1% respondents strongly agreed to understand what it means to be a responsible digital citizen; 41.4% respondents agreed and 40% respondents strongly agreed on likes to learning while using digital tools.

**Figure 6. Reading Enjoyment and Preference**

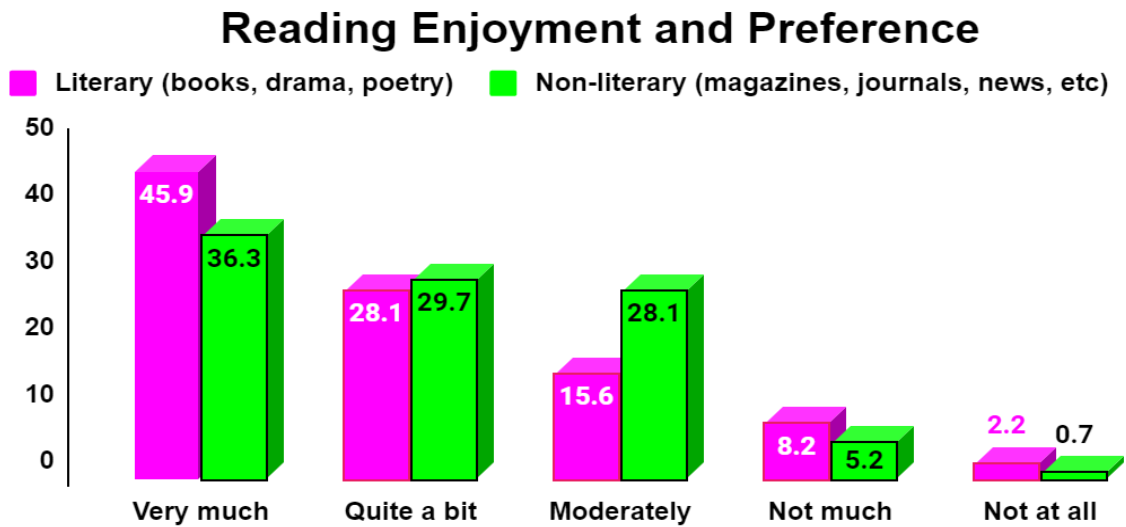

Figure 6 shows that reading enjoyable literary books or not literary books. As a result, in the above table 45.9% of respondents liked literary books very much, 36.3% of respondents liked non-literary books very much.

**Figure 7. Reasons for Enjoying Reading**

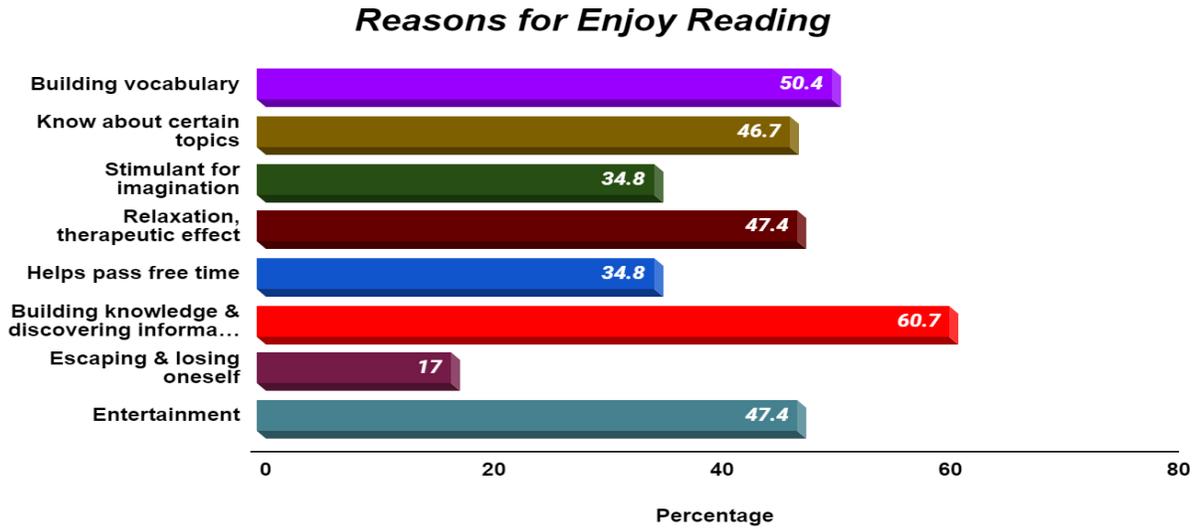

Considering the result in figure 7 reasons for enjoying reading. Most of the respondents building knowledge, discovering new information 60.7%.

**Figure 8. Reading recommendations received from**

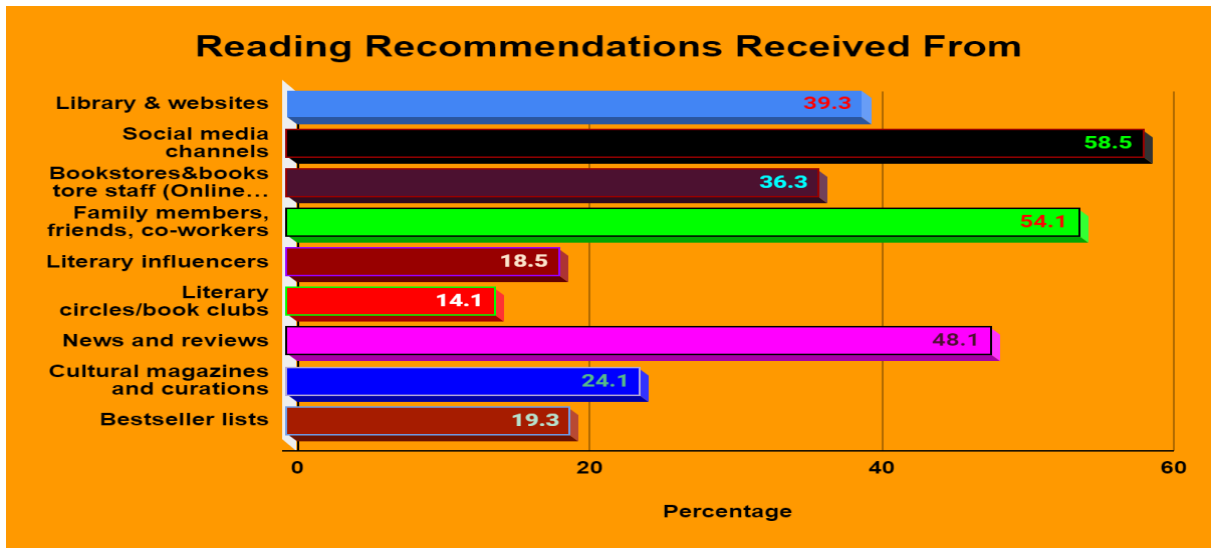

As depicted in figure 8, students and scholars received a source of reading recommendations. 58.5% of respondents received reading recommendations from social media channels, followed by 54.1% respondents from family members, friends, and coworkers, 48.1% respondents from news and reviews, and 14.1% the least respondents received from literary circles and book reviews.

**Figure 9. Format types and priority resource for reading**

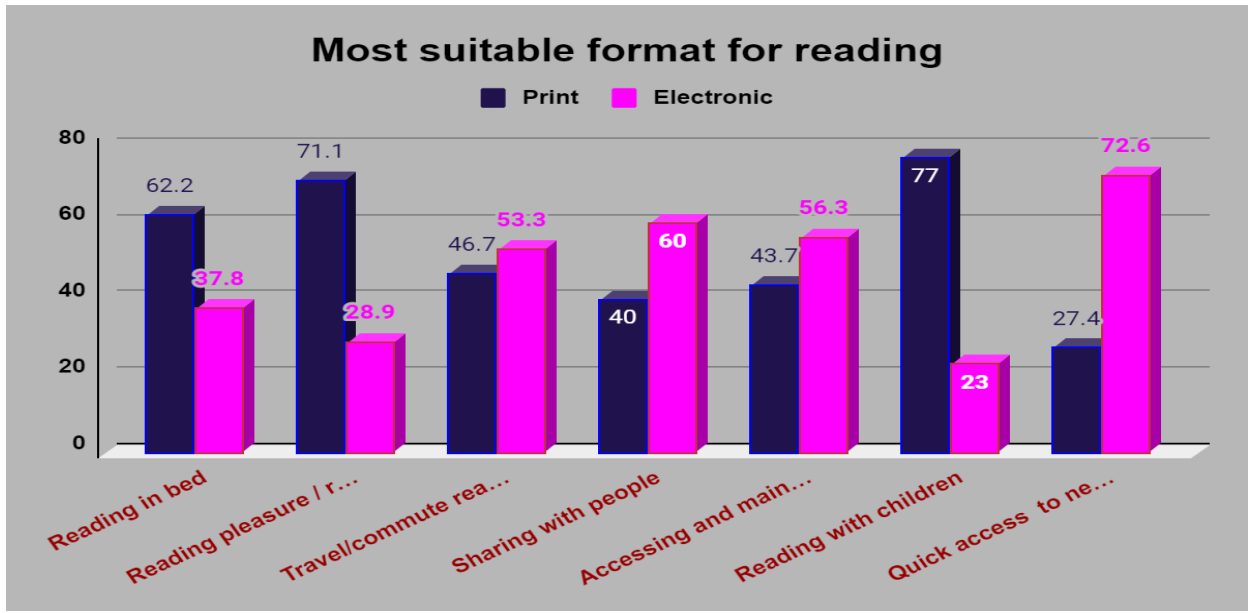

Figure 9 shows electronic or print books which are suitable for reading. Reading in bed majority of them preferred print 62.2% respondent, reading for pleasure/recreational value print format is most of the option 71.1% of respondents, Travel/commute reading suitable is the electronic format preferred by 53.3% of respondents, sharing with people appropriate format preferred is electronic 60% respondents, accessing and maintaining a wide collection of books applicable format is electronic 56.3% respondents, reading with children convenient format is print 77% of respondents, quick access to new material adaptable format is electronic 72.6% of respondents.

**Figure 10. Digital Media and Information Literacy is one of the ways to Paperless Society**

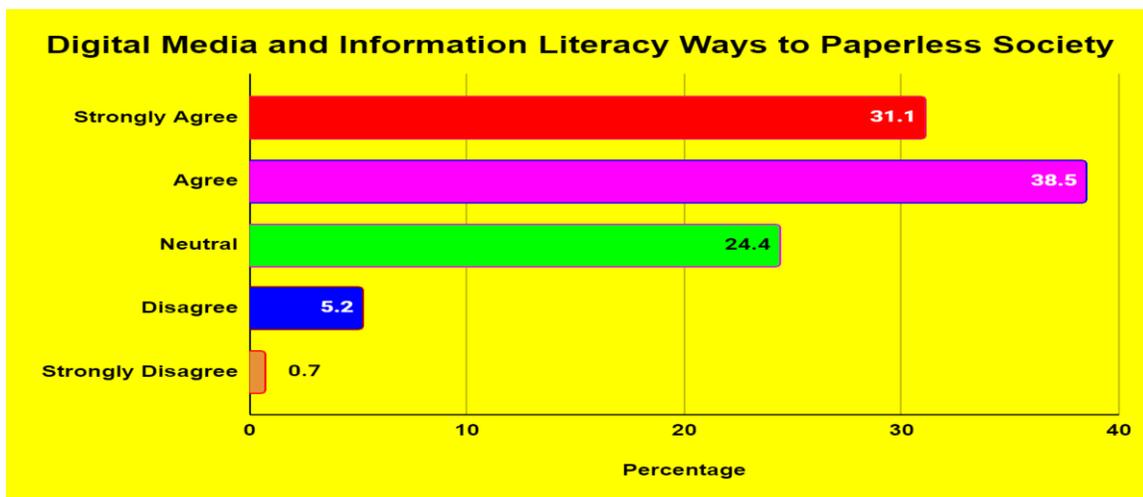

As illustrated in figure 10 that digital media and information literacy skills are a way to a paperless society. For this statement 38.5% of respondents agreed, 31.1% of respondents strongly agreed, 24.4% of respondents were neutral, 5.2% of respondents disagreed and 0.7% of respondents very least people only strongly disagreed.

### 7. Conclusion

Digital literacy and digital-based reading are most important in the present scenario. Libraries have various kinds of resources such as primary, secondary and territory and multiple formats of resources are print, electronic, multimedia. This study proved that digital literacy is reached everywhere whether it's urban or rural. That's not at all a matter. Technology emerged and students also adopted ICT for their day-to-day life. For in-depth knowledge reading habits are prominent at the same time digital literacy skills are too important for evaluating the resources. Misinformation and disinformation are spreading everywhere like a virus for that understanding the virus digital literacy is working as an anti-virus. Students willing to read whatever the source is print or digital but are highly comfortable with digital media-based reading. Digital and media literacy is one of the ways to reach the paperless society and library because when one person is educated, they know how to access the technology and use the technology.

**References**
1. Anjaiah, M. (2016). Digital Information Literacy Among Research Scholars and Students Community At Dravidian University, Kuppam-Andra Pradesh (India): An Exploratory Study. *IOSR Journal of Humanities and Social Science*, *21*(09), 01–08. https://doi.org/10.9790/0837-2109120108
2. Ayhan, B. (2016). Digital Literacy. In *Digitalization and Society* (pp. 29–48). Peter Lang. https://doi.org/10.3726/978-3-653-07022-4/10
3. Ben Amram, S., Aharony, N., & Bar Ilan, J. (2021). Information Literacy Education in Primary Schools: A Case Study. *Journal of Librarianship and Information Science*, *53*(2), 349–364. https://doi.org/10.1177/0961000620938132
4. Emiri, O. T. (2017). Digital Literacy Skills Among Librarians in University Libraries In the 21st Century in Edo And Delta States, Nigeria. *International Journal of Library and Information Services*, *6*(1), 37–52. https://doi.org/10.4018/IJLIS.2017010103
5. Erciyes University, Yamaç, A., Erciyes University, & Öztürk, E. (2019). How Digital Reading Differs from Traditional Reading: An Action Research. *International Journal of Progressive Education*, *15*(3), 207–222. https://doi.org/10.29329/ijpe.2019.193.15
6. Goodwin, A. P., Cho, S.-J., Reynolds, D., Brady, K., & Salas, J. (2020). Digital Versus Paper Reading Processes and Links to Comprehension for Middle School Students. *American Educational Research Journal*, *57*(4), 1837–1867. https://doi.org/10.3102/0002831219890300

7. Iqbal Singh, B. (2015). Digital Information Literacy among Health Sciences Professionals: A Case Study of GGS Medical College, Faridkot, Punjab, India. *Proceedings of Informing Science & IT Education Conference (InSITE) 2015*, 531–541. https://www.informingscience.org/Publications/2149?Source=%2FConferences%2FInSITE2015%2FProceedings
8. Jeffrey, L., Hegarty, B., Kelly, O., Penman, M., Coburn, D., & McDonald, J. (2011). Developing Digital Information Literacy in Higher Education: Obstacles and Supports. *Journal of Information Technology Education: Research*, *10*, 383–413. https://doi.org/10.28945/1532
9. Johnston, N. (2020). The Shift towards Digital Literacy in Australian University Libraries: Developing a Digital Literacy Framework. *Journal of the Australian Library and Information Association*, *69*(1), 93–101. https://doi.org/10.1080/24750158.2020.1712638
10. Khatun, M., Virkus, S., & Rahman, A. I. M. J. (2015). Digital Information Literacy: A Case Study in Oslo Public Library. In S. Kurbanoglu, J. Boustany, S. Špiranec, E. Grassian, D. Mizrachi, & L. Roy (Eds.), *Information Literacy: Moving Toward Sustainability* (Vol. 552, pp. 121–131). Springer International Publishing. https://doi.org/10.1007/978-3-319-28197-1_13
11. Khosrow-Pour, M. (Ed.). (2017). *Encyclopedia of Information Science and Technology: 10* (4th edition). IGI Global.
12. Lim, F. V., & Toh, W. (2020). How to Teach Digital Reading? *Journal of Information Literacy*, *14*(2), 24–43. https://doi.org/10.11645/14.2.2701
13. McDougall, J., Readman, M., & Wilkinson, P. (2018). The Uses of (digital) Literacy. *Learning, Media and Technology*, *43*(3), 263–279. https://doi.org/10.1080/17439884.2018.1462206
14. Medlock Paul, C., Spires, H., & Kerkhoff, S. (2017). Digital Literacy for the 21st Century. In *Encyclopedia of Information Science and Technology* (pp. 2235–2242). IGI-Global. https://doi.org/10.4018/978-1-5225-7659-4.ch002
15. Nelson, K., Courier, M., & Joseph, G. (2011). An Investigation of Digital Literacy Needs of Students. *Journal of Information Systems Education*, *22*(2), 95–110.
16. Parvathamma, N., & Pattar, D. (2013). Digital literacy among student community in management institutes in Davanagere District, Karnataka State, India. *Annals of Library and Information Studies (ALIS)*, *60*(3), 159–166.
17. Shopova, T. (2014). Digital Literacy of Students and Its Improvement at the University. *Journal on Efficiency and Responsibility in Education and Science*, *7*(2), 26–32. https://doi.org/10.7160/eriesj.2014.070201
18. Shwetha, K. K., & Mallaiah, T. Y. (2017). Digital Information Literacy Skills Among Faculty Members of Engineering Colleges in Managalore, Karantaka: A Study. *International Journal of Digital Library Services*, *7*(1), 28–37.

19. Techataweewan, W., & Prasertsin, U. (2018). Development of Digital Literacy Indicators for Thai Undergraduate Students Using Mixed Method Research. *Kasetsart Journal of Social Sciences*, *39*(2), 215–221. https://doi.org/10.1016/j.kjss.2017.07.001
20. The Faculty of Education, Chulalongkorn University, Bangkok, Thailand, Kaeophanuek, S., Jaitip, N.-S., & Nilsook, P. (2018). How to Enhance Digital Literacy Skills among Information Sciences Students. *International Journal of Information and Education Technology*, *8*(4), 292–297. https://doi.org/10.18178/ijiet.2018.8.4.1050